\documentclass[a4paper,fleqn,usenatbib]{mnras}
\usepackage[draft]{hyperref}

\usepackage[T1]{fontenc}
\usepackage{ae,aecompl}

\usepackage{graphicx}	
\usepackage{amsmath}	
\usepackage{amssymb}	

\title[Gaia 1 and 2]{Gaia 1 and 2. A pair of new Galactic star clusters}

\author[S. Koposov et al.]{
Sergey E. Koposov,$^{1,2}$\thanks{E-mail: skoposov@cmu.edu}
V. Belokurov,$^{1}$
G. Torrealba$^{1}$
\\
$^{1}$Institute of Astronomy, University of Cambridge, CB3 0HA\\
$^{2}$McWilliams Center for Cosmology, Department of Physics, Carnegie
Mellon University, 5000 Forbes Avenue, Pittsburgh, PA 15213, USA
}
\date{Accepted XXX. Received YYY; in original form ZZZ}

\pubyear{2017}

\begin{document}
\label{firstpage}
\pagerange{\pageref{firstpage}--\pageref{lastpage}}
\maketitle

\begin{abstract}

We present the results of the very first search for faint Milky Way
satellites in the {\it Gaia} data. Using stellar positions only, we are able
to re-discover objects detected in much deeper data as recently as the
last couple of years. While we do not identify new prominent
ultra-faint dwarf galaxies, we report the discovery of two new star
clusters, Gaia~1 and Gaia~2. Gaia~1 is particularly curious, as it is
a massive (2.2$\times$10$^4$ M$_\odot$), large ($\sim$9 pc) and nearby
(4.6\, kpc) cluster, situated 10$\arcmin$ away from the brightest star
on the sky, Sirius! Even though this satellite is detected at
significance in excess of 10, it was missed by previous sky
surveys. We conclude that {\it Gaia} possesses powerful and unique
capabilities for satellite detection thanks to its unrivalled angular
resolution and highly efficient object classification.
\end{abstract}

\begin{keywords}
Galaxy: general -- Galaxy: globular clusters -- Galaxy: open clusters and
associations -- catalogues -- Galaxy: structure
\end{keywords}



\section{Introduction}

In 1785, William Herschel introduced ``star gauging'' as a technique
to understand the shape of the Galaxy. Herschel suggested that by
splitting the heavens into cells and by counting the number of stars
in each, it should be possible to determine the position of the Sun
within the Milky Way. Even though he promoted the method as ``the most
general and the most proper'', Herschel was quick to point out the
possibility of strong systematic effects given that different
telescopes would probe the skies to different depths. Over the
centuries, Herschel's prophecy of the merit of the technique was
confirmed, and most recently, the data from the all-sky surveys such
as  Two Micron All-Sky Survey (2MASS) \citep{2mass} and Sloan Digital Sky Survey (SDSS) \citep{york00} has been used to show off
its true power. However, as the telescope-might grew and yet fainter
stars were counted, the problem of minute but systematic depth
variations blossomed into a major hindrance to studies of the Milky
Way structure.

\begin{figure*}
\includegraphics[]{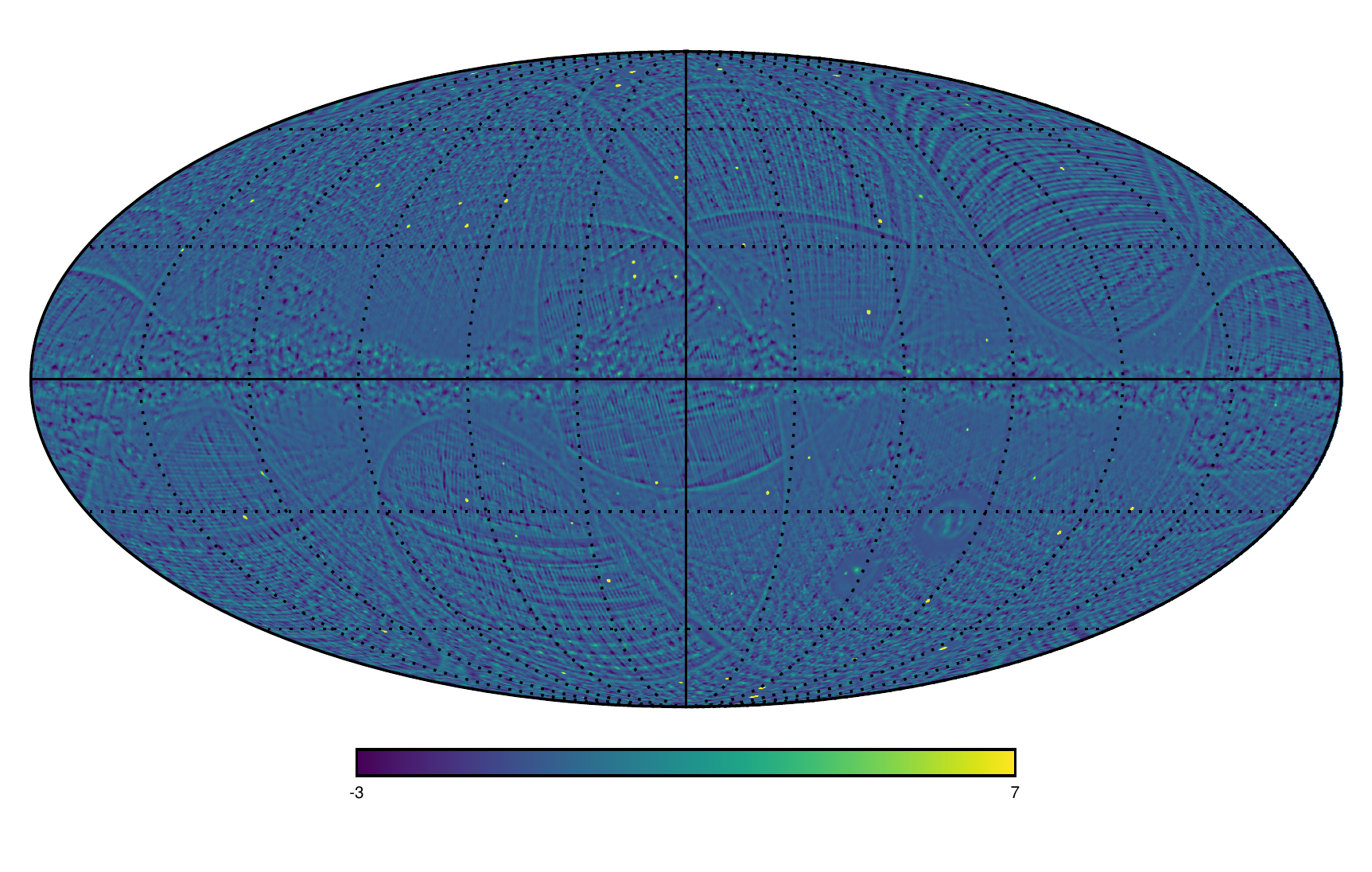}
\caption{The all-sky map of significances of stellar overdensities in 
Mollweide projection. The map was obtained using a 12$\arcmin$ Gaussian kernel.}
\label{fig:sign_map}
\end{figure*}

In the Galaxy, the one application that suffers the most from the
non-uniform quality of the ``star gage'' is the search for the faint
sub-structures in the stellar halo. Ultra-faint satellites and stellar
streams appear as low-level enhancements of the stellar density field
and as such can either be destroyed or mimicked by the survey
systematics. For example, for a given exposure on a given telescope, a
combination of the atmosphere stability, the sky brightness and the
image resolution controls the limiting magnitude at which stars can be
distinguished from galaxies with certainty. Of these three factors
above, only the camera properties are mostly invariant with time, while - as
seen from the ground - the sky is constantly changing. Thus, the faint
star counts will fluctuate as a function of the position on the sky in
accordance with the survey progress. Around the limiting magnitude,
the loss of genuine stars due to poor weather conditions is exacerbated by
the influx of spurious ``stars'', i.e. compact faint galaxies
misclassified as stellar objects. At low Galactic latitudes,
diffraction spikes, blooming and ghosts produced by bright stars also
contribute large numbers of fictitious ``stars''. As a result,
typically, over-densities of bogus stellar objects caused by
misclassified galaxies in galaxy clusters and by artifacts around
bright stars outnumber bona-fide satellites by several orders of
magnitude.

Today, some 230 years after Herschel's first attempt, {\it Gaia}, the
European Space Astrometric Mission, as part of the Data Release 1 (DR1), has
produced the most precise Star Gage ever \citep{perryman01,gaia_mission16,gaia_dr1}. Positions,
magnitudes and a wealth of additional information on over a billion sources, as recorded in
the {\tt GaiaSource} table, do not suffer from poor weather conditions
or dramatic sky brightness variations. Better still, any spurious
objects reported by the on-board detection algorithm, are
double-checked and discarded after subsequent visits. In terms of the
star-galaxy separation, {\it Gaia} is truly unique. Its angular
resolution is only approximately a factor of two worse than the resolution of the 
WFC3 camera on-board of the HST and {\it Gaia} can additionally discriminate between stars and galaxies
based on the differences in their astrometric behaviour and spectrophotometry. Motivated by
the unique properties of the {\it Gaia} DR1 data, here we investigate the
mission's capabilities for satellite detection. The Paper is organised
as follows. Section~\ref{sec:detection} presents the details of the
satellite detection algorithm as applied to the {\it Gaia} DR1 data. The
{\it Gaia} discoveries are discussed in
Section~\ref{sec:discovery}. Section~\ref{sec:properties} studies the
properties of the new satellites and Section~\ref{sec:discussion} concludes the paper.

\section{Satellite detection with {\it Gaia} DR1}
\label{sec:detection}

\subsection{{\tt GaiaSource} all sky catalogue}

The entirety of the analysis presented in this paper is based on the
first {\it Gaia} data product released in late 2016, the {\tt {GaiaSource}}  table. For details on the contents and the properties of
this catalogue we refer the reader to the {\it Gaia} Data Release 1 paper
\citep[][]{gaia_dr1}. Note that in this paper the only quantities from the
catalogue that we use are the 
stellar positions (RA and Dec), the G-band magnitudes \citep{gaia_phot} and 
the
values of the {\tt astrometric\_excess\_noise} parameter
\citep{lindegren16}.

\subsection{Satellite search algorithm}

The principal ideas behind our Galactic satellite detection algorithm
have been extensively covered in the literature \citep[see
  e.g.][]{irwin94, belokurov07,
  koposov08a,koposov08b,walsh09,willman2010,koposov15,bechtol15}. However,
the {\it Gaia} DR1 data represents a rather unusual and - in parts -
challenging dataset to apply these simply ``out of the box''. This is
primarily due to the absence of the object colour information, unless,
of course, cross-matched with other surveys, which are usually either
less deep than {\it Gaia}, such as 2MASS and APASS or do not
cover the whole sky. Additionally, at magnitudes fainter than $G=20$,
the survey's depth starts to vary significantly as a function of the
position on the sky. As a result, in this very early data release, the
{\it Gaia} sky appears to look like an intricate pattern of gaps of varying
sizes \citep[see e.g.][]{gaia_dr1}.

The combination of the factors above drove us to introduce a slight
modification to the stellar overdensity identification method. We do
use the kernel density estimation with Gaussian kernels of varying
width, however in order to assess the significance of the density
deviation at any given point on the sky, we do not rely on the Poisson
distribution of the stellar number counts. Instead we obtain a local
estimate of the variance of the density. More precisely, if $K(. ,.)$
is the properly normalised density kernel on the sphere,
$H(\alpha,\delta)$ is the histogram of the star counts \citep[on
  HEALPix grid on the sky,][]{gorski05}, the density estimate is
simply the convolution $D(\alpha,\delta) = K \ast H$ and the
significance of the overdensity $S(\alpha,\delta) = \frac { D -
  <D>_A}{\sqrt{Var_A(D)}}$ , where the average $<D>_A$ and the
variance $Var_A(D)$ are calculated over the HEALPix neighbourhood
(annulus) of a given point. Given the normalisation by local variance,
areas with pronounced non-uniformities related to the {\it Gaia} scanning
law, or regions with large changes in extinction are naturally
down-weighted in their contribution to significance.

Given that the first {\it Gaia} DR does not provide colour information, we
have to be particularly careful when selecting objects for the
satellite search \citep[we cannot use colour-magnitude masks based on
  stellar isochrones, as in i.e.][]{koposov15}. For the analysis in
this paper we apply only two selection cuts. First cut concerns the
{\it Gaia} $G$ magnitude: $17<G<21$. The reason for getting rid of bright
stars, i.e. those with $G<17$ is twofold. First, for absolute majority
of interesting targets at reasonable distances from the Sun we expect
the luminosity function of their stellar populations to rapidly rise
with magnitude at $G>17$. Therefore, in these satellites (globular
clusters and dwarf galaxies), the number of stars fainter than $G=17$
should vastly overwhelm the number of bright stars. Instead, the
brighter magnitudes are dominated by the foreground
contamination. Equally important is the fact that most (or all)
satellites with significant populations of stars at $G<17$ have likely
been already detected, either through studies of the 2MASS photometry
\citep[e.g.][]{koposov08b} or via the inspection of the archival
photographic plate data \citep[see e.g.][]{whiting07}.

The cut at faint magnitudes $G<21$ is less straightforward. Usually,
when dealing with large optical ground-based surveys, it is advisable
to restrict the data to be limited by the apparent magnitude which
gives the most uniform density map. However, in our case, due to the
relatively shallow {\it Gaia} DR1 depth in combination with the rising
stellar luminosity function for objects of interest, the benefits of
going to e.g. $G=21$ as compared to $G=20.5$ outweigh the drawbacks of
dealing with non-uniform density distribution, caused by spatially
varying incompleteness \citep[see e.g.][]{gaia_dr1}. Accordingly, for
our analysis we set the faint limit at $G=21$.

The second and final cut applied to the catalogue concerns the
star/galaxy separation. While the {\it Gaia}'s on-board detection algorithm
was designed with point sources in mind, the observatory nonetheless
detects (and reports) compact galaxies at faint magnitudes as well as
compact portions of extended galaxies. In fact, in the high latitude
(|b|>30$\degr$) area where {\it Gaia} overlaps with the SDSS, at magnitude $G=20$
around $10$\% of sources detected by {\it Gaia} and reported in the {\tt
  GaiaSource} catalogue are classified by the SDSS as galaxies (bear in
mind though that the SDSS star-galaxy classification is not 100\%
correct). As galaxies are much more clustered than stars, it is
crucial to filter them out from the sample before embarking on the
search for over-densities. In the {\tt GaiaSource} catalogue, the
parameter which correlates the most with the object's non-stellarity
is the {\tt astrometric\_excess\_noise} metric which measures the
extra scatter in the astrometric solution of all objects
\citep[][]{lindegren16}. Here, in order to reject extended sources in
the magnitude range $17<G<21$ we adopt the following
magnitude-dependent cut on the {\tt astrometric\_excess\_noise}
parameter:
\begin{equation}\log_{10}({\tt   astrometric\_excess\_noise}) < 0.15\, (G-15) + 0.25 \label{eq:star_gal}\end{equation}
The quality of this selection can be assessed using the SDSS catalogue:
in the magnitude range $19<G<20$, more than 95\% of {\it Gaia} sources
rejected by the above cut are classified by SDSS as galaxies
confirming that this is indeed a very effective way to weed out
spurious ``stars'' from the {\it Gaia} stellar sample.

\begin{figure}
\includegraphics[]{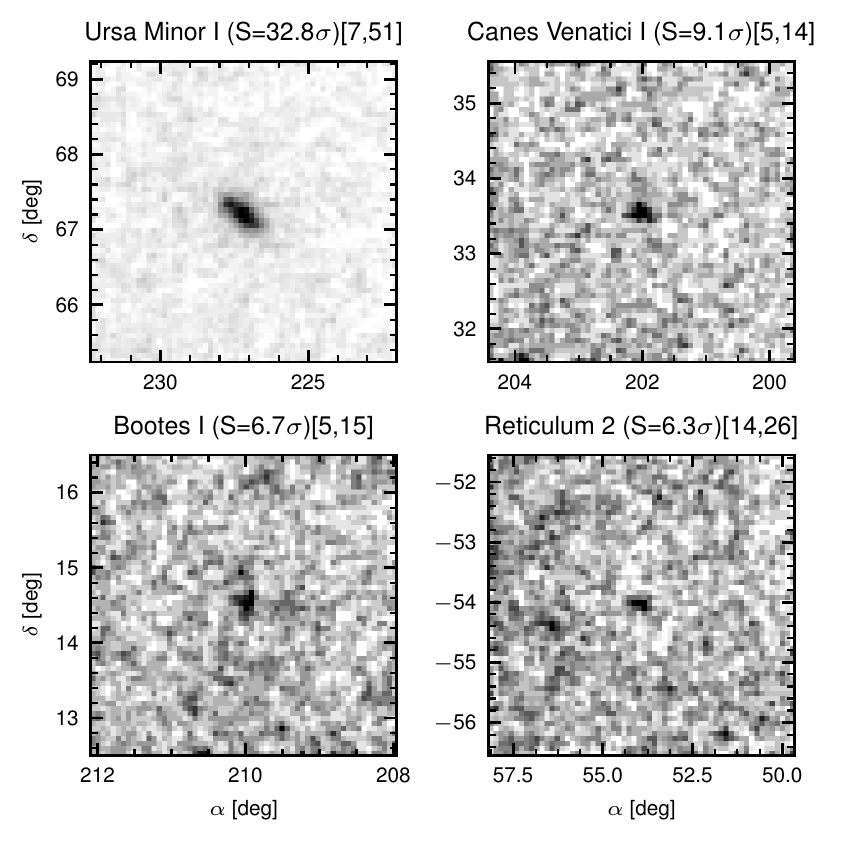}

\caption{The gallery of known faint and ultra-faint Milky Way satellites.
The greyscale shows the density of {\it Gaia} sources with $17<G<21$ and
satisfying the stellarity selection (Eq.~\ref{eq:star_gal}). The significance of the
detection of each object is shown in the title. The number of stars
per bin corresponding to white and black colour in the density maps is given in
brackets in the panel titles.}
\label{fig:knowns}
\end{figure}

Figure~\ref{fig:sign_map} shows the map of significances of 
overdensities as measured using the Gaussian kernel with $\sigma=
12\arcmin$. This map is produced using the entire {\tt
  GaiaSource} catalogue filtered with the cuts described above. Note the
compact yellow spots, the vast majority of which correspond to the
overdensities associated with the known Galactic satellites. The
underlying fine web of green-blue filaments is related to the {\it Gaia}
scanning law.

\section{Detected Objects}
\label{sec:discovery}

\begin{figure*}
\includegraphics[]{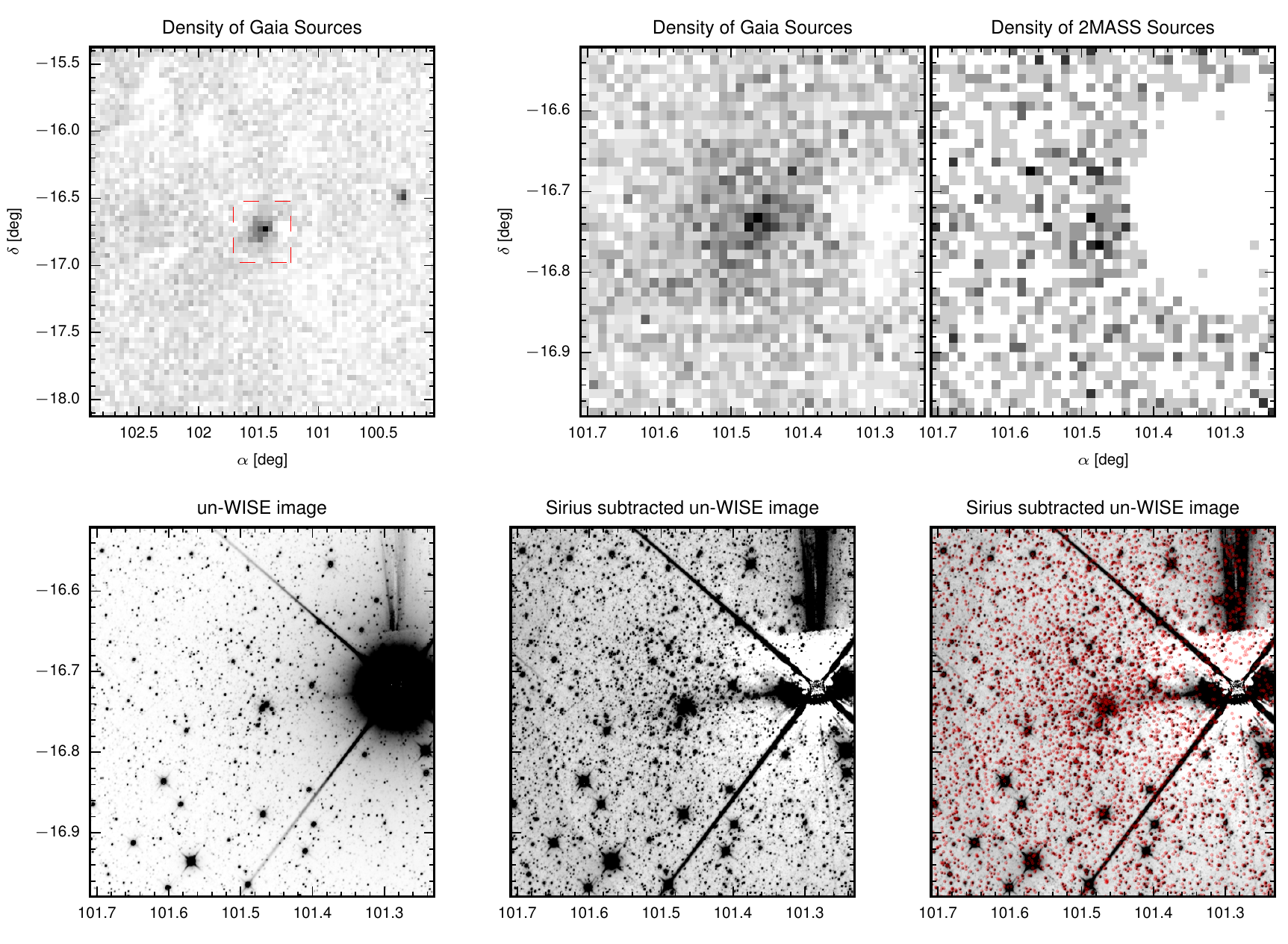}
\caption{Distribution of sources around Gaia~1. {\it Top left:} The
  density of all {\it Gaia} sources with $G<20$ within $\sim$ 3 degrees of
  Gaia~1. The overdensity on the right edge of the panel is
  Berkeley~25, an old open cluster. The red box indicates the size of
  the region shown on other panels of the figure. {\it Top middle:} The
  density of {\it Gaia} sources with $G<20$ in the $\sim 30\arcmin\times 30
  \arcmin$ field of view around Gaia~1. {\it Top right:} The density
  of the 2MASS sources in the same field of view. {\it Bottom left:}
  The $30\arcmin\times 30 \arcmin$ image from the WISE survey showing
  Gaia~1. {\it Bottom middle:} The same image with the PSF of Sirius
  subtracted. {\it Bottom right:} The same image with {\it Gaia} source
  positions overplotted in red.}
\label{fig:gaia_2ddensity}
\end{figure*}

\subsection{Known objects}

In this paper, we choose to focus only on the most obvious detections,
i.e. those above 6$\sigma$ in significance. From the suite of runs of
the overdensity detection algorithm with different kernels ranging
from 1.5$\arcmin$ to 24$\arcmin$, a total of 259 such candidates were
identified. At the next step, these were cross-matched with various
catalogues of star clusters and nearby galaxies
\citep{dias02,kontizas90,mccon12,harris10,makarov14} as well as the
CDS's Simbad database \citep{wenger00}. Out of the 259 objects, 244
had a clear association with a known star cluster or a galaxy. To our
surprise the list of previously known satellites detected in the {\it Gaia}
DR1 data at high significance was not limited to the classical
globular clusters and the classical dwarf galaxies, known for
decades. Amongst the detections, we found several ultra-faint dwarf
galaxies, such as Canes Venatici I \citep{zucker06}, detected at
significance of 9.1 $\sigma$, Bootes I \citep{belokurov06} detected at
6.7 $\sigma$, and Reticulum 2 \citep{koposov15,bechtol15} with
significance of 6.3. Some other, fainter dwarfs such as Canes
Venatici~II \citep{belokurov07} and Crater~2 \citep{torrealba16} were
also detected, albeit at slightly lower significance (below our
nominal threshold), i.e. 5.8 and 5.1 respectively.

\begin{table}
\caption{Measured parameters of clusters detected in Gaia dataset}

\begin{tabular}{ccc}
\hline
Name & Gaia~1 & Gaia~2\\
\hline
Right Ascension [deg]: & $101.47$ & $28.124 \pm 0.007$  \\ 
Declination [deg]: & $-16.75$ & $53.040 \pm 0.005$ \\ 
Galactic longitude [deg] & $202.34$ &  $131.909$ \\
Galactic latitude [deg] & $-8.75$ & $-7.764$ \\
Half-light radius[$\arcmin$]: & $6.5 \pm 0.4$
& $1.90^{+0.4}_{-0.34}$ \\ 
Half-light radius[pc]: & $9 \pm 0.05$
& $3^{+0.63}_{-0.53}$ \\ 
Ellipticity: & $-$ & $0.18^{+0.2}_{-0.12}$ \\ 
Distance modulus (m-M): & $13.3\pm0.1$ & $13.65\pm0.1$ \\
Age [Gyr]\footnotemark: & $6.3\pm 1$ & $8 \pm 2$\\
Absolute magnitude (V): & $-5 \pm 0.1$ & $-2 \pm 0.1$\\
\hline
\end{tabular}
\label{tab:params}
\end{table}
\footnotetext{Due to limited colour magnitude information, we consider ages
highly uncertain}

Figure~\ref{fig:knowns} shows the density distributions of 
stellar sources around some of the dwarf galaxies
detected in {\it Gaia} DR1. These range from one of the most prominent
dwarfs, namely UMi, to the notoriously difficult to find ultra-faints,
several of which were detected less than two years ago in the Dark Energy
Survey data \citep{koposov15,bechtol15}. These examples demonstrate
the incredible purity and quality of the {\tt GaiaSource} catalogue, and
highlight {\it Gaia}'s superb satellite discovery capabilities even without colour information.

While exploring the candidate over-density list, we have stumbled upon
two previously unknown objects detected with very high significance,
which are going to be the focus of the rest of the paper.

\subsection{Gaia~1. You can not be Sirius!}

The first stellar overdensity without an obvious counterpart,
i.e. without an entry in any of the clusters/galaxies catalogues
available to us, has an estimated significance of $\sim$ 10. Note
however, that initially, this candidate, dubbed here Gaia~1, was
rejected as it is located in close proximity of the brightest star on
the sky -- Sirius\footnote{Previously when working with ground-based
  surveys such as SDSS, VST and DES the authors usually automatically
  excluded the regions around bright stars as CCD saturation, ghost
  reflections and diffraction spikes produced by bright stars usually
  lead to spurious overdensities in the
  catalogues.}. Figure~\ref{fig:gaia_2ddensity} shows the distribution
of sources in the area around the candidate from three different
surveys: {\it Gaia}, 2MASS and WISE \citep{wright2010}. As the Figure
demonstrates, the overdensity in {\it Gaia}-detected sources is indeed very
prominent (top left and top middle panels). Note a white patch
corresponding to a dearth of {\it Gaia} sources near the centre of the
over-density. This is the location of Sirius, where {\it Gaia} struggles to
detect genuine stars\footnote{It is curious that in the {\it Gaia} paper on
  source list creation by \citet{fabricius16} the region around Sirius
  was used to demonstrate challenges of dealing with such a bright
  source (see their Fig. 12)}. The area affected by Sirius is much
larger in the 2MASS data, as evidenced in the right panel. However,
even here, a stellar overdensity corresponding to the candidate in
question is noticeable.  The object itself can be seen directly in the
images from the infrared WISE survey (bottom row). The bottom panels
of Figure~\ref{fig:gaia_2ddensity} show the view of the $\sim
30\arcmin\times 30\arcmin$ area around the overdensity as provided by
the un-WISE project \citep{lang14}. The left panel displays the
original WISE data with Sirius in the picture, while the middle panel
has the star subtracted using a circularly symmetric PSF model.  In
both cases the overdensity of sources in the central parts of Gaia~1
is clearly visible. This is the most straightforward and the most
secure confirmation of the newly identified satellite.

\begin{figure} \includegraphics[]{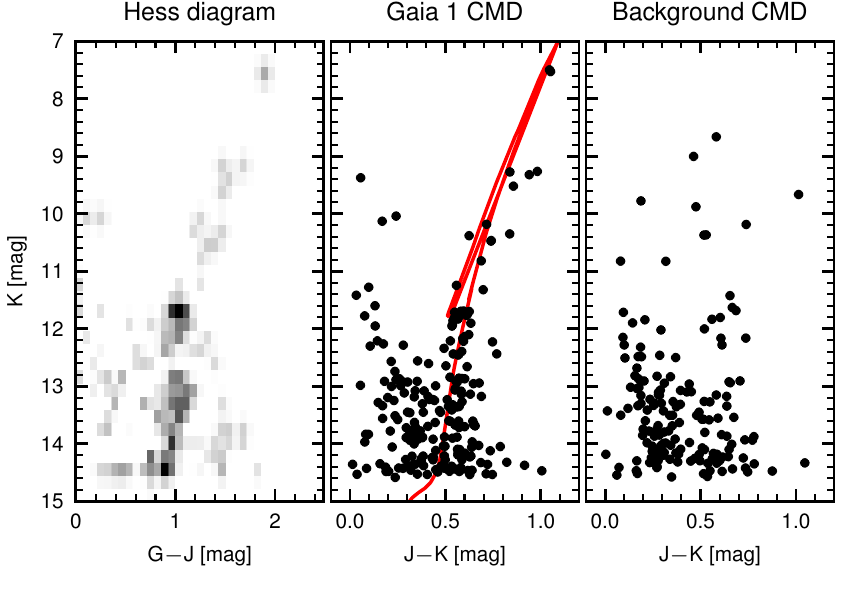} \caption{{\it Left panel:}
The background subtracted and extinction corrected 2MASS-{\it Gaia} Hess diagram of the central
0.1\degr\ of Gaia~1 (area within the annulus with inner and outer radii of
0.3 and 1 degrees has been used for background); {\it Middle panel:} The
2MASS $J-K_s,K_s$ colour-magnitude diagram of Gaia~1 obtained using sources
within $0.1$\,deg from Gaia~1 centre.  The PARSEC isochrone with the age of 
$6.3$\,Gyr and ${\rm [Fe/H]}=-0.7$ at the distance modulus of $m-M=13.3$ is
overplotted in red.  {\it Right panel:} The colour-magnitude diagram of the
background field offset by $0.5$\degr\ from Gaia~1 and with the same area as
used for the middle panel.} \label{fig:gaia1_cmd} \end{figure}
\begin{figure} \includegraphics[]{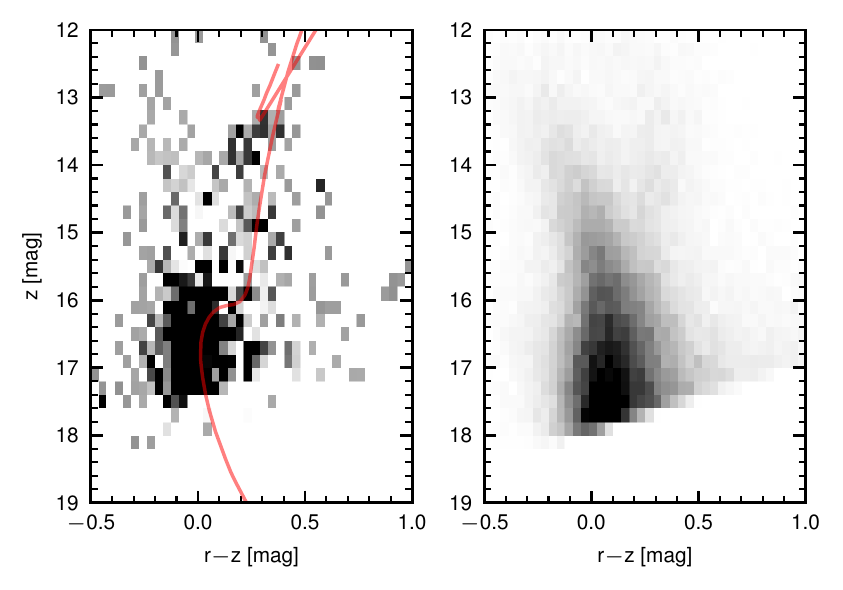} \caption{{\it Left panel:}
The background subtracted $r-z$, $z$ Hess diagram of the central 0.1\degr\ of
Gaia~1 from PS1 data.  The PARSEC isochrone with the age of $6.3$\,Gyr and ${\rm
[Fe/H]}=-0.7$ at the distance modulus of $m-M=13.3$ is overplotted in red. 
{\it Right panel:} The Hess diagram of the stellar foreground in the
0.3\degr, 1\degr annulus around Gaia~1.} \label{fig:gaia1_cmd_ps}
\end{figure}

Our next step is to try to assess the nature of the overdensity by
means of studying its colour-magnitude diagram
properties. Unfortunately, the {\it Gaia} DR1 release does not contain colour
information, so we are forced to use much shallower 2MASS data. Left
panel of Figure~\ref{fig:gaia1_cmd} shows the extinction-corrected
Hess diagram obtained by combining 2MASS and {\it Gaia} photometry for the
field centred on the Gaia~1 (to get the combined 2MASS/{\it Gaia} 
photometry we used the nearest neighbour cross-match between surveys with 
3$\arcsec$ aperture). The red giant branch together with
prominent red clump (or red horizontal branch) at $K_s\sim 11.7$ are
both clearly visible. The other two panels of the Figure show the
2MASS-only colour magnitude diagrams (CMD).  The middle panel gives the CMD
of the stars within the central parts of Gaia~1 together with the
PARSEC isochrone \citep{bressan12} with ${\rm[Fe/H]}=-0.7$ and age of
$6.3$\,Gyr. The right panel of the Figure displays the
colour-magnitude distribution of the foreground stars in the part of
the sky $0.5$ degrees away from the centre of Gaia~1 with the
same area as the field shown in the middle panel. We note that the
isochrone reproduces the location of the red clump and the red giant
branch and that the colour-magnitude diagram distribution in Gaia~1 is
clearly different from the distribution of the foreground stars, which
provides another confirmation that the object is a genuine satellite.

A better idea of the object's stellar populations can be gleaned from
the recently released Pan-STARRS1 (PS1) data
\citep{ps1_main,ps1_data,ps1_cats}. Due to the presence of an
extremely bright star nearby, the quality of the PS1 data in the
region is significantly compromised but the catalogues remain immensely
useful. We obtain a list of sources around Gaia~1 from the stack detections
catalogue (the {\tt StackObjectThin} table) and require that the sources
have the following flags: STACK\_PRIMARY, PSFMODEL, FITTED on, and do
not have the SKY\_FAILURE flag. This together with the cut on
extinction-corrected r-band magnitude of $r<17.5$ provides a
relatively clean source subset and allows us to finally peer at the
deeper colour-magnitude diagram of the satellite. Left panel of
Figure~\ref{fig:gaia1_cmd_ps} shows the $r-z$ vs $z$
 extinction corrected \citep{schlafly11} background subtracted Hess diagram, while the Hess diagram of the
foreground stellar population is displayed on the right. With the PS1
data being deeper compared to 2MASS, we can see the large number of
turn-off stars at $r-z\sim 0$ and $z\sim16$. Note, that these are
exactly the stars that contributed most of the signal in the {\it Gaia}
overdensity detection. The PS1 data also shows the RHB (or the red
clump) at $r-z\sim$ 0.3. As the Figure indicates, the $6.3$\,Gyr and
${\rm [Fe/H]}=-0.7$ isochrone also provides a decent description of
the PS1 data.

\begin{figure*}
\includegraphics[]{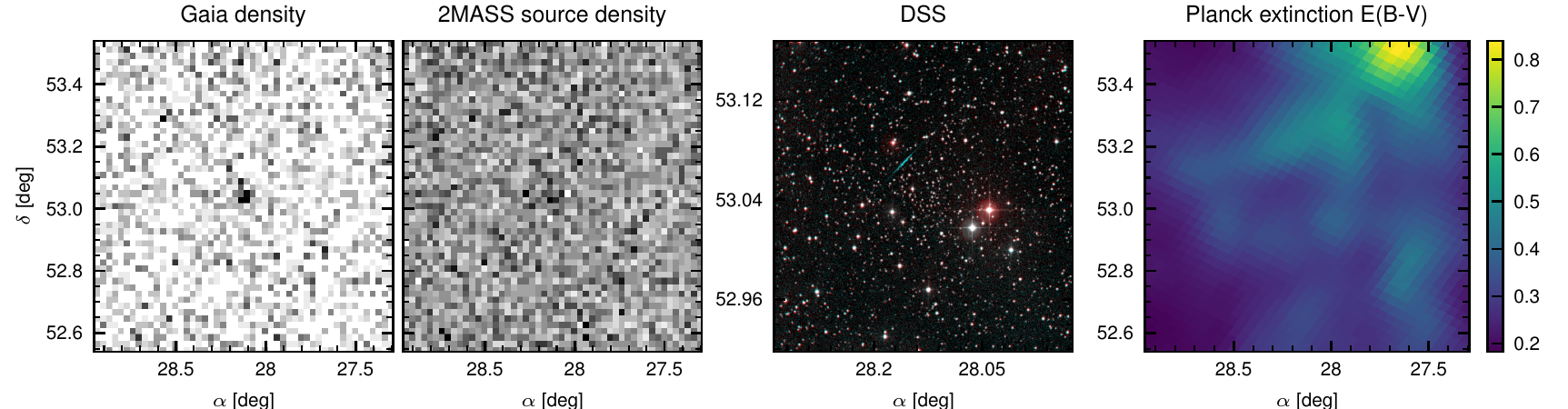}
\caption{{\it Left panel:} The density of {\it Gaia} sources with extinction
corrected {\it Gaia} magnitude brighter than 19 in the $1\degr \times 1\degr$
area around Gaia~2. 
{\it Second panel:} The density of 2MASS sources around Gaia~2. 
{\it Third panel:} The colour DSS image of $15\arcmin \times 15\arcmin$\ field of view around Gaia~2. {\it Fourth panel:} The extinction map around Gaia~2.}
\label{fig:gaia2}
\end{figure*}

\subsection{Gaia~2}

The second object in our list possesses a significance of
9$\sigma$. This overdensity is located in dense stellar field of the
sky, close to the galactic plane at $b=-8.7\degr$.  Figure~\ref{fig:gaia2}
shows an array of discovery plots confirming that the candidate is
indeed a {\it bona-fide} satellite. The left panel of the Figure gives
the density distribution of the {\it Gaia} sources with $G_0<19$ (where
$G_0$ is the extinction-corrected {\it Gaia} G-magnitude) and $\tt
astrometric\_excess\_noise<5$. Here we use the \citet{sfd} dust map
and the extinction coefficient of 2.55 \citep[see][]{belokurov16}. It
is clear that {\it Gaia} has detected a pronounced stellar overdensity in
the centre of the field. The next panel gives the density distribution
of the 2MASS sources and here, similarly to Gaia~1 we see an
overdensity at lower significance. The middle right panel displays the
colour mosaic of the Digital Sky Survey (DSS) (Blue/Red) images of the area. Note an
agglomeration of stars clearly visible in the centre of the image. The
rightmost panel presents the dust extinction distribution as reported
by the Planck satellite \citep[see][]{Planck14}. While some spatial
variation of extinction is noticeable in the area, including a small
decrease in reddening in the very centre of the field, there exists no
correlation between the dust distribution and the {\it Gaia} stellar density
distribution. Thus, based on the evidence listed above, we conclude
that this overdensity is also a genuine Galactic satellite, named
Gaia~2 hereafter.

\begin{figure}
\includegraphics[]{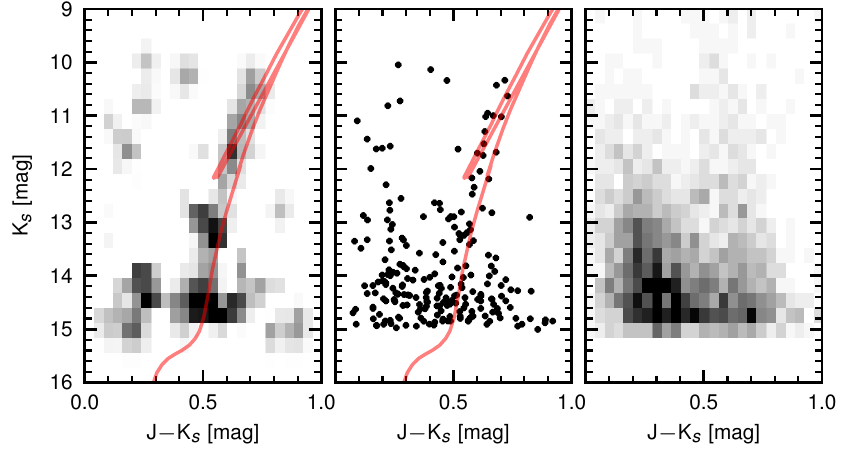}
\caption{{\it Left panel:} The background subtracted and extinction
corrected 2MASS Hess diagram
  of Gaia~2 within central 3\arcmin. The PARSEC isochrone with the age of 8\,Gyr and [Fe/H]$=-0.6$ at a distance modulus of 14.65 is overlaid in red. {\it Middle panel:} The 
	2MASS colour-magnitude diagram of Gaia~2
  within 3\arcmin\ of centre. {Right panel:} The Hess diagram of the
  background stars in the 12\arcmin, 24\arcmin\ annulus.}
\label{fig:gaia2_cmd}
\end{figure}

\begin{figure}
\includegraphics[]{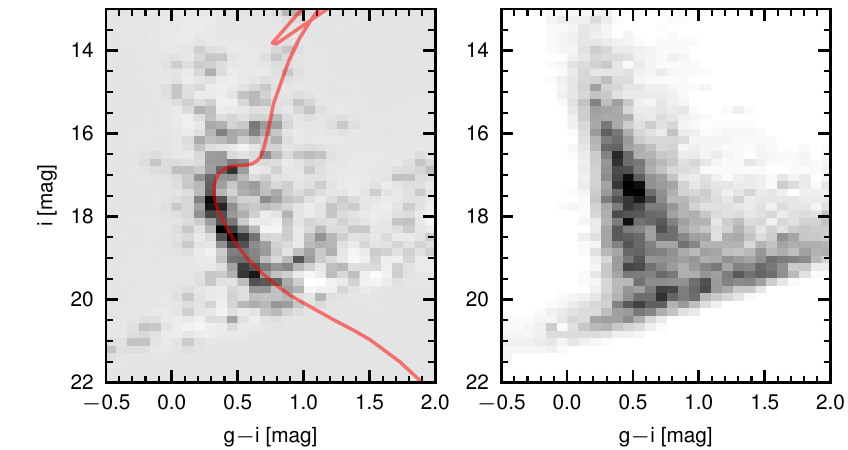}
\caption{{\it Left panel:} The background subtracted and extinction
corrected PS1 Hess
  diagram of Gaia~2 within central 3\arcmin. The PARSEC isochrone with the age of 8\,Gyr and [Fe/H]$=-0.6$ at a distance modulus of 14.65 is overlaid in red. {Right panel:} 
The Hess diagram of the background stars in the 12\arcmin, 24\arcmin\
annulus.}
\label{fig:gaia2_cmd_ps1}
\end{figure}

Figure~\ref{fig:gaia2_cmd} gives a glimpse of the colour-magnitude
distribution of the stars in Gaia~2. The 2MASS-based Hess diagram
(left) assuredly exhibits the red-giant branch with a possible red
clump.  Note that both of these CMD features are clearly absent in the
foreground population (right panel of the Figure). For Gaia~2, we also
utilise the PS1 data
to obtain a deeper colour-magnitude
diagram of the object. Accordingly, Figure~\ref{fig:gaia2_cmd_ps1}
shows the background subtracted and extinction corrected Hess diagram
of the central parts of the Gaia~2 in the PS1 data, while the right
panel displays the background Hess diagram; both panels use the $g$
and $i$ bands. The depth of the PS1 data is sufficient to see the main
sequence (MS) of the satellite. The presence of the obvious MS is the
final confirmation of the nature of this stellar overdensity.


\section{Properties of Gaia~1 and 2}
\label{sec:properties}
\subsection{Gaia~1}

\begin{figure}
\includegraphics[]{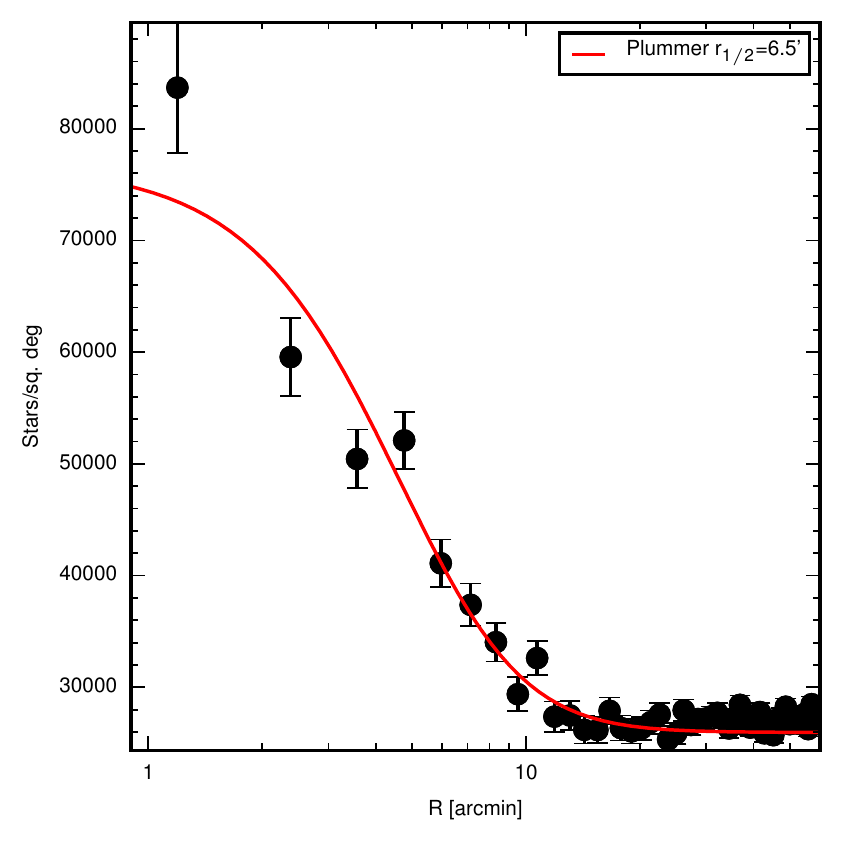}
\caption{The 1D density profile of Gaia~1. The red line shows a Plummer model fit with half-light radius of $6.5\arcmin$. We also note that in the very centre there is a tight group of stars which causes the innermost point in the density profile to deviate from Plummer.}
\label{fig:gaia1_1dprof}
\end{figure}
The extinction-corrected photometry of the satellite as supplied by
2MASS and PS1 is relatively shallow, thus limiting our ability
to accurately measure the system's age, metallicity and
distance. However the problem is remedied to some degree by the
presence of the obvious Red Clump, visible in the colour-magnitude
diagram discussed above, that allows robust distance determination.
The 2MASS extinction corrected K magnitude of the Red Clump of Gaia~1
is $K_s = 11.7$ (with a statistical uncertainty less than 0.01
mag). Although the absolute magnitude of the Red Clump might depend
weakly on the age and the metallicity of the stellar population, we
assume $M_K=-1.6$ \citep[see][and discussion
  there]{williams2013,girardi16}. This gives us a distance modulus of
$m-M\sim 13.3\pm 0.1$ corresponding to $\sim 4.6\pm0.2$ kpc (assuming a systematic uncertainty of $\sim$ 0.1 on the red-clump absolute
magnitude). With the distance estimate in hand we can now constrain
the age and the metallicity of the system by fitting isochrones to the
data. Here we do not try to carry out the full stellar population
modelling, due to the limitations of the data currently available. We
simply point out that the $6.3$\,Gyr and ${\rm [Fe/H]}=-0.7$ isochrone
at the distance dictated by the red clump magnitude appears to describe
well both the 2MASS red giant branch and the main sequence turn-off
visible in the PS1 data (see Figures~\ref{fig:gaia1_cmd} and
\ref{fig:gaia1_cmd_ps}). We estimate that the age uncertainty is 
$\sim$ 1$-$2\, Gyr and the uncertainty in ${\rm [Fe/H]}$ is $\sim$ 0.2 dex.

As a next step we proceed with the determination of structural parameters.
Due to the proximity of Sirius and possible spatially
variable incompleteness in the catalogue at faint magnitudes, we have
decided to use only stars with $G<19$ and located far enough from
Sirius ($\alpha>101.4\degr$). For the same reason, instead of doing
the 2-D density analysis we perform an un-binned 1-D density profile
fitting using the Plummer model for the satellite and the uniform
background density. Table~\ref{tab:params} gives the results of the
fit together with their uncertainties as determined from 1-D
posteriors, which were sampled using the {\tt emcee} ensemble sampler
\citep{emcee}. Figure~\ref{fig:gaia1_1dprof} shows the Gaia~1's
observed 1D density profile together with our best-fit Plummer model.
We note that innermost point of the density profile is deviating
slightly from the model, and that the 2-D stellar distribution in
Gaia~1 seems to have a rather compact group of stars in the very
centre (see top middle panel of Figure~\ref{fig:gaia_2ddensity}). We
are not able, at this point, to determine the exact cause of this
over-density, or even to rule out an artefact unambiguously. The
satellite's half-light radius determined from the fit is
$6.5\arcmin\pm 0.4$ which corresponds to $\sim$ 9\, pc. Based on the
size typical for globular clusters in the MW,
we conclude that Gaia~1 is a star cluster, most likely of globular
variety given its age and metallicity. It also could be an old open
cluster similar to e.g. Berkeley~25 located only $\sim$1 degree away
on the sky from Gaia~1.  From the density profile fit we also
determine the total number of the satellite's stars with $G$<19,
$N\sim 1200\pm 120$. Assuming the Chabrier IMF \citep{chabrier03} and the best-fit
isochrone, we deduce the total stellar mass of 22000 $M_\odot$ and the
$V$-band luminosity of $M_V\sim -5 \pm 0.1$. Note though that these numbers
are only ball-park estimates due to the uncertain age and
metallicity of the stellar population.
 
To investigate the cluster's orbital properties, it would be interesting
to gauge the proper motion of the Gaia~1's stellar
members. Unfortunately, due to the proximity to Sirius, the bright
Gaia~1 members are not present in the TGAS catalogue, so in order to 
make a proper measurement we likely have to wait for the second release of 
Gaia data.

\subsection{Gaia~2}

The properties of the Gaia~2's stellar population, namely its
metallicity and age can be deduced from the combination of the 2MASS
and PS1 photometry. However, because the PS1 data is
limited to the lower Main Sequence, this analysis is subject to a
substantial age-metallicity degeneracy. Through experiments with
PARSEC isochrones, we find that the stellar population with 8\,Gyr
and ${\rm [Fe/H]}=-0.6$ fits both the lower Main Sequence in PS1
and the Red Giant Branch in 2MASS. We estimate the age uncertainty to 
be $\sim$ $1-2$\,Gyr 
and metallicity uncertainty $\sim$ 0.2 dex. The distance modulus implied by
this isochrone is $m-M=13.65\pm 0.1$ corresponding to the distance of
$5.4$\,kpc. A better imaging is definitely required to make a more
detailed determination of the age, metallicity and distance of this
object.

To measure the structural parameters of Gaia~2 we model the
distribution of the {\tt GaiaSource} number counts. We restrict the
sample to stars with $G_0<19$. The spatial density distribution is
then modelled with a combination of a flat background and an elliptical
Plummer profile for the satellite \citep[for
more details see e.g.][]{koposov15}. The resulting measurements are given in the
Table~\ref{tab:params}. We find that the object is mildly elliptical,
although at low significance levels. The half-light radius of the
object is $\sim 1.9$\arcmin, which at the distance of $5.4$\,kpc
corresponds to $3$\,pc. Thus, Gaia~2 is almost three times smaller compared
to Gaia~1. The total luminosity of the cluster can be estimated
similarly to that of Gaia~1 using the best-fit isochrone and the total
number of stars with $G_0<19$, resulting in $M_V=-2\pm 0.1$. Given its size,
metallicity, age and appearance on the image cutouts, it is most
likely a globular cluster, or an old open cluster.
 
\section{Conclusions}
\label{sec:discussion}

This paper presents the results of the very first exploration of the
{\it Gaia} capabilities for Galactic satellite detection. While we have
identified a number of new interesting satellite candidates, here we
describe only two objects, Gaia~1 and Gaia~2, whose nature can be
determined unambiguously with extant data. Possessing remarkably high
levels of significance - in excess of 9 (!) - both were, nevertheless,
missed by previous imaging surveys. Thus, {\it Gaia} appears to offer a
unique view of the Galactic stellar distribution, unrivalled even by
datasets reaching to significantly fainter magnitudes. It is due to
the combination of the {\it Gaia}'s high angular resolution and the high
purity of its stellar sample, objects like Gaia~1 can be discovered.

Gaia~1 is a large and luminous star cluster. However, it is positioned almost
exactly behind Sirius, the brightest star in the sky, which appears to
be the perfect hiding place even for a satellite of such grand
prominence. Countless generations of astronomers must have stared
right at Gaia~1, blissfully unaware of its existence. The area around
Sirius is difficult to analyse as it is littered with artifacts
spawned by this extremely bright star. However, given the {\it Gaia}'s
multi-epoch data-stream, such artifacts can be weeded out in the
preparation of the {\tt GaiaSource} catalogue.  Each time {\it Gaia} looks at
Sirius, the spurious sources appear in different positions on the sky
compared to the previous visits and thus are mostly discarded
due to the DR1 requirement of having 5 or more detections of a single source
\citep{lindegren16}. Even if some spurious sources make it to the final {\tt
  GaiaSource} catalogue, they can be identified very efficiently based
on their unusual astrometric behaviour.

We have demonstrated that the {\it Gaia} data can be used to detect
classical dwarf galaxies, ultra faint dwarfs and star clusters in a
wide range of sizes and luminosities. All of these are detected with
ease using the {\it Gaia}'s stellar positions only in the catalogue which is
known to suffer from spatially-variable incompleteness. We therefore
predict with confidence, that in the future {\it Gaia} Data Releases, when
the completeness is stabilised and the colour, distance and proper
motion information is added, many new galactic satellites will be
revealed.

\section*{Acknowledgements}

We acknowledge the usage of the HyperLeda database
(http://leda.univ-lyon1.fr).  This research has made use of the SIMBAD
database, operated at CDS, Strasbourg, France.  The research leading
to these results has received funding from the European Research
Council under the European Union's Seventh Framework Programme
(FP/2007-2013)/ERC Grant Agreement no. 308024. SK thanks the United
Kingdom Science and Technology Council (STFC) for the award of Ernest
Rutherford fellowship (grant number ST/N004493/1). SK also thanks Bernie Shao for the assistance
with obtaining the Pan-STARRS1 data. The authors thank the anonymous referee for careful reading of 
the manuscript.

This paper was written in part at the 2016 NYC Gaia Sprint, hosted by the Center for Computational Astrophysics at the Simons Foundation in New York City.

This research made use of {\tt Astropy}, a community-developed core Python package
for Astronomy \citep{astropy} and {\tt Q3C} extension for {\tt PostgreSQL} database \citep{koposov06}.

This work has made use of data from the European Space Agency (ESA)
mission {\it Gaia} (\url{http://www.cosmos.esa.int/gaia}), processed by
the {\it Gaia} Data Processing and Analysis Consortium (DPAC,
\url{http://www.cosmos.esa.int/web/gaia/dpac/consortium}). Funding
for the DPAC has been provided by national institutions, in particular
the institutions participating in the {\it Gaia} Multilateral Agreement.

The Pan-STARRS1 Surveys (PS1) and the PS1 public science archive have been made possible through contributions by the Institute for Astronomy, the University of Hawaii, the Pan-STARRS Project Office, the Max-Planck Society and its participating institutes, the Max Planck Institute for Astronomy, Heidelberg and the Max Planck Institute for Extraterrestrial Physics, Garching, The Johns Hopkins University, Durham University, the University of Edinburgh, the Queen's University Belfast, the Harvard-Smithsonian Center for Astrophysics, the Las Cumbres Observatory Global Telescope Network Incorporated, the National Central University of Taiwan, the Space Telescope Science Institute, the National Aeronautics and Space Administration under Grant No. NNX08AR22G issued through the Planetary Science Division of the NASA Science Mission Directorate, the National Science Foundation Grant No. AST-1238877, the University of Maryland, Eotvos Lorand University (ELTE), the Los Alamos National Laboratory, and the Gordon and Betty Moore Foundation.

\bsp	
\label{lastpage}
\end{document}